\def\be{\begin{equation}}
\def\ee{\end{equation}}
\def\ba{\begin{array}}
\def\ea{\end{array}}

\documentclass[aps,amsmath,amssymb,amsfonts,showpacs]{revtex4}
\usepackage{graphicx}
\usepackage{caption2}
\usepackage{epstopdf}
\def\qed{\leavevmode\unskip\penalty9999 \hbox{}\nobreak\hfill
     \quad\hbox{\leavevmode  \hbox to.77778em{%
               \hfil\vrule   \vbox to.675em%
               {\hrule width.6em\vfil\hrule}\vrule\hfil}}
     \par\vskip3pt}

\newtheorem{theorem}{Theorem}

\newtheorem{lemma}{Lemma}

\begin{document}
\title{ General Monogamy Relations of Quantum Entanglement for Multiqubit W-class States}
\author{Xue-Na Zhu$^{1}$}
\author{Shao-Ming Fei$^{2,3}$}

\affiliation{$^1$School of Mathematics and Statistics Science, Ludong University, Yantai 264025, China\\
$^2$School of Mathematical Sciences, Capital Normal University,
Beijing 100048, China\\
$^3$Max-Planck-Institute for Mathematics in the Sciences, 04103
Leipzig, Germany}

\begin{abstract}

Entanglement monogamy is a fundamental property of multipartite
entangled states. We investigate the monogamy relations for multiqubit generalized W-class states.
Analytical monogamy inequalities are obtained for the concurrence of assistance,
the entanglement of formation and the entanglement of assistance.

\end{abstract}

\maketitle

\section{Introduction}

Quantum entanglement \cite{t1,t2,t3,t4,t5,t6} is an essential feature of
quantum mechanics that distinguishes the quantum from
the classical world. It is one of the fundamental differences
between quantum entanglement and classical correlations that
a quantum system entangled with one of the other systems
limits its entanglement with the remaining others. This restriction of
entanglement shareability among multi-party systems is
known as the monogamy of entanglement. The monogamy relations give rise
to the structures of entanglement in the multipartite setting.
For a tripartite system A, B, and C, the monogamy of
an entanglement measure $\varepsilon$ implies that the entanglement
between $A$ and $BC$ satisfies $\varepsilon_{A|BC}\geq\varepsilon_{AB}+
\varepsilon_{AC}$.

In Ref.\cite{JSK1,JSK2} the monogamy of entanglement for multiqubit $W$-class states has been investigated, and
the monogamy relations for tangle and the squared concurrence have been proved.
In this paper, we show the general monogamy relations for the $x$-power of concurrence of assistance,
the entanglement of formation,
and the entanglement of assistance for generalized multiqubit $W$-class states.

\section{Monogamy of concurrence of assistance}

For a bipartite pure state $|\psi\rangle_{AB}$ in vector space $H_A\otimes H_{B}$,
the concurrence is given by \cite{c1,c2,c3}
\begin{equation}\label{CON}
C(|\psi\rangle_{AB})=\sqrt{2[1-Tr(\rho^2_A)]},
\end{equation}
where $\rho_A$ is reduced density matrix by tracing over the subsystem $B$,
$\rho_{A}=Tr_{B}(|\psi\rangle_{AB}\langle\psi|)$.
The concurrence is extended to mixed states $\rho=\sum_{i}p_{i}|\psi _{i}\rangle \langle \psi _{i}|$,
$p_{i}\geq 0$, $\sum_{i}p_{i}=1$, by the convex roof construction,
\begin{equation}\label{CONC}
C(\rho_{AB})=\min_{\{p_i,|\psi_i\rangle\}} \sum_i p_i C(|\psi_i\rangle),
\end{equation}
where the minimum is taken over all possible pure state decompositions of $\rho_{AB}$.

For a tripartite state $|\psi\rangle_{ABC}$, the concurrence of assistance (CoA) is defined by \cite{ca}
\begin{equation}
C_a(|\psi\rangle_{ABC})\equiv C_a(\rho_{AB})
=\max_{\{p_i,|\psi_i\rangle\}}\sum_ip_iC(|\psi_i\rangle),
\end{equation}
for all possible ensemble realizations of
$\rho_{AB}=Tr_{C}(|\psi\rangle_{ABC}\langle\psi|)=\sum_i p_i |\psi_i\rangle_{AB} \langle \psi_i|$.
When $\rho_{AB}=|\psi\rangle_{AB}\langle \psi|$ is a pure state, then one has
$C(|\psi\rangle_{AB})=C_{a}(\rho_{AB})$.

For an $N$-qubit state $|\psi\rangle_{AB_1...B_{N-1}}\in H_A\otimes H_{B_1}\otimes...\otimes H_{B_{N-1}}$, the concurrence
$C(|\psi\rangle_{A|B_1...B_{N-1}})$ of the state
$|\psi\rangle_{A|B_1...B_{N-1}}$, viewed as a bipartite
with partitions $A$ and $B_1B_2...B_{N-1}$, satisfies the follow
inequality\cite{024304}
\begin{equation}\label{cm1}
C^\alpha_{A|B_1B_2...B_{N-1}}\geq C^\alpha_{AB_1}+C^\alpha_{AB_2}+...+C^\alpha_{AB_{N-1}},
\end{equation}
and
\begin{equation}\label{cm2}
C^\beta_{A|B_1B_2...B_{N-1}}< C^\beta_{AB_1}+C^\beta_{AB_2}+...+C^\beta_{AB_{N-1}},
\end{equation}
where $\alpha\geq2$, $\beta\leq0$, $C_{AB_i}=C(\rho_{AB_i})$ is the concurrence of $\rho_{AB_i}=Tr_{B_1...B_{i-1}B_{i+1}...B_{N-1}}(\rho)$,
$C_{A|B_1B_2...B_{N-1}}=C(|\psi\rangle_{A|B_1...B_{N-1}})$.
Due  to the monogamy of concurrence,
the generalized monogamy relation
based on the concurrence of assistance has been proved in Ref. \cite{dualmonogamy},
\begin{equation}
C^2(|\psi\rangle_{A|B_1...B_{N-1}})\leq \sum_{i=1}^{N-1}C^2_a(\rho_{AB_i}).
\end{equation}

In the following we study the monogamy property of the concurrence of assistance for the $n$-qubit
generalized W-class states $|\psi\rangle\in H_{A_1}\otimes H_{A_2}\otimes...\otimes H_{A_n}$ defined by
\begin{equation}\label{w}
|\psi\rangle=a|000...\rangle+b_1|01...0\rangle+...+b_n|00...1\rangle,
\end{equation}
with $|a|^2+\sum_{i=1}^{n}|b_i|^2=1$.

\begin{lemma}\label{c=ca}
For $n$-qubit generalized W-class states (\ref{w}), we have
\begin{eqnarray}\label{cca}
C(\rho_{A_1A_i})=C_a(\rho_{A_1A_i}),
\end{eqnarray}
where $\rho_{A_1A_i}=Tr_{A_2...A_{i-1}A_{i+1}...A_{n}}(|\psi\rangle\langle\psi|)$.
\end{lemma}

[Proof] It is direct to verify that \cite{JSK1},
$\rho_{A_1A_i}=|x\rangle_{A_1A_i}\langle x|+|y\rangle_{A_1A_i}\langle y|$,
where
$$
\ba{l}
|x\rangle_{A_1A_{i}}=a|00\rangle_{A_1A_{i}} +b_1|10\rangle_{A_1A_{i}}+b_i|01\rangle_{A_1A_{i}},\\[2mm]
|y\rangle_{A_1A_{i}}=\sqrt{\sum_{k\neq i}|b_k|^2}|00\rangle_{A_1A_{i}}.
\ea
$$
From the Hughston- Jozsa-wootters theorem Ref.\cite{JSK1}, for any pure-state decomposition of
$\rho_{A_1A_{i}}=\sum_{h=1}^{r}|\phi_h\rangle_{A_1A_{i}}\langle\phi_h|$, one has
$|\phi_h\rangle_{A_1A_{i}} =u_{h1}|x\rangle_{A_1A_{i}}+u_{h2}|y\rangle_{A_1A_{i}}$ for
some $r\times r$ unitary matrices $u_{h1}$ and $u_{h2}$ for each $h$.
Consider the normalized state $|\tilde{\phi_h}\rangle_{A_1A_{i}}=|\phi_h\rangle_{A_1A_{i}}/\sqrt{p_h}$ with $p_h=|\langle\phi_h|\phi_h\rangle|$.
One has the concurrence of each two-qubit pure $|\tilde{\phi_h}\rangle_{A_1A_{i}}$,
\begin{equation}\nonumber
C^2(|\tilde{\phi_h}\rangle_{A_1A_{i}})=\frac{4}{p_h^2}|u_{hi}|^4|b_1|^2|b_i|^2.
\end{equation}
Then for the two-qubit state $\rho_{A_1A_{i}}$, we have
\begin{equation}\nonumber
\sum_hp_hC(|\tilde{\phi_h}\rangle_{A_1A_{i}})
=\sum_hp_h\frac{2}{p_h}|u_{hi}|^2|b_1||b_i|
=2|b_1||b_i|.
\end{equation}
Thus we obtain
\begin{eqnarray}\nonumber
C(\rho_{A_1A_i})
&=&
\min_{\{p_h,|\tilde{\phi_h}\rangle_{A_1A_{i}}\}}
\sum_hp_hC(|\tilde{\phi_h}\rangle_{A_1A_{i}})\\[1mm]\nonumber
&=&\max_{\{p_h,|\tilde{\phi_h}\rangle_{A_1A_{i}}\}}
\sum_hp_hC(|\tilde{\phi_h}\rangle_{A_1A_{i}})\\[1mm]\nonumber
&=&C_a(\rho_{A_1A_i}).
\end{eqnarray}
\qed

Specifically, in Ref. \cite{JSK2} the same result $C(\rho_{A_1A_i})=C_a(\rho_{A_1A_i})$ has been proved
for the generalized W-class states (\ref{w}) with $a=0$.

\begin{theorem}\label{TH2}
For the $n$-qubit generalized W-class states $|\psi\rangle\in H_{A_1}\otimes H_{A_2}\otimes...\otimes H_{A_n}$, the
concurrence of assistance satisfies
\begin{eqnarray}\label{cax}
C_a^x(\rho_{A_1|A_{j_1}...A_{j_{m-1}}})\geq\sum_{i=1}^{{m-1}}C^x_a(\rho_{A_1A_{j_i}}),
\end{eqnarray}
where $x\geq2$ and $\rho_{A_1A_{j_1}...A_{j_{m-1}}}$ is the $m$-qubit, $2\leq m\leq n$, reduced density matrix of $|\psi\rangle$.
\end{theorem}

[Proof]
For the $n$-qubit generalized W-class state $|\psi\rangle$,
according to the definitions of $C(\rho)$ and $C_a(\rho)$, one has
$C_a(\rho_{A_1|A_{j_1}...A_{j_{m-1}}})\geq C(\rho_{A_1|A_{j_1}...A_{j_{m-1}}})$. When $x\geq2$, we have
\begin{eqnarray}\nonumber
C^x_a(\rho_{A_1|A_{j_1}...A_{j_{m-1}}})
&\geq&C^x(\rho_{A_1|A_{j_1}...A_{j_{m-1}}})\\[1mm]\nonumber
&\geq&\sum_{i=1}^{m-1}C^x(\rho_{A_1A_{j_i}})\\[1mm]\nonumber
&=& \sum_{i=1}^{m-1}C^x_a(\rho_{A_1A_{j_i}}).
\end{eqnarray}
Here we have used in the first inequality the inequality $a^x\geq b^x$ for $a\geq b>0$ and $x\geq0$.
The second inequality is due to the monogamy of concurrence (\ref{cm1}).
The last equality is due to the Lemma \ref{c=ca}.

\begin{theorem}\label{TH3}
For the $n$-qubit generalized W-class state $|\psi\rangle\in H_{A_1}\otimes H_{A_2}\otimes...\otimes H_{A_n}$
with $C(\rho_{A1A_{j_i}})\neq0$ for $1\leq i\leq m-1$, we have
\begin{eqnarray}\label{cay}
C_a^y(\rho_{A_1|A_{j_1}...A_{j_{m-1}}})<\sum_{i=1}^{{m-1}}C^y_a(\rho_{A_1A_{j_i}}),
\end{eqnarray}
where $y\leq0$ and $\rho_{A_1A_{j_1}...A_{j_{m-1}}}$ is the $m$-qubit reduced density matrix as in Theorem 1.
\end{theorem}

[Proof] For $y\leq0$, we have
\begin{eqnarray}\nonumber
C^y_a(\rho_{A_1|A_{j_1}...A_{j_{m-1}}})
&\leq &C^y(\rho_{A_1|A_{j_1}...A_{j_{m-1}}})\\[1mm]\nonumber
&< &\sum_{i=1}^{m-1}C^y(\rho_{A_1A_{j_i}})\\[1mm]\nonumber
&=& \sum_{i=1}^{m-1}C^y_a(\rho_{A_1A_{j_i}}).
\end{eqnarray}
We have used in the first inequality the relation $a^x\leq b^x$ for $a\geq b>0$ and $x\leq0$.
The seconder inequality is due to the monogamy of concurrence (\ref{cm2}). The last equality is due to Lemma 1.

According to (\ref{cax}) and (\ref{cay}), we can also obtain the lower bounds of $C_a(\rho_{A_1|A_{j_1}...A_{j_{m-1}}})$.
As an example, consider the $5$-qubit generalized $W$-class states (\ref{w}) with
$a=b_2=\frac{1}{\sqrt{10}}$, $b_1=\frac{1}{\sqrt{15}}$, $b_3=\sqrt{\frac{2}{15}}$, $b_4=\sqrt{\frac{3}{5}}$.
We have
$$C_a(\rho_{A_1|A_{2}A_{3}})\geq\frac{2}{\sqrt{15}}\sqrt[x]{(\frac{1}{\sqrt{10}})^x
+(\sqrt{\frac{2}{15}})^x}
$$
and
$$
C_a(\rho_{A_1|A_{2}A_{3}A_{4}})
\geq\frac{2}{\sqrt{15}}\sqrt[x]{(\frac{1}{\sqrt{10}})^x
+(\sqrt{\frac{2}{15}})^x+\sqrt{\frac{3}{5}})^x}
$$
with $x\geq2$. The optimal lower bounds can be obtained by varying the parameter $x$,
see Fig. 1, where for comparison the upper bounds are also presented by using the
formula $C_a(\rho_{AB})\leq\sqrt{2(1-Tr(\rho_{A}^2))}$ \cite{ZGL}, namely,
$C_a(\rho_{A_1|A_{2}A_{3}})\leq\frac{2}{\sqrt{18}}$
and $C_a(\rho_{A_1|A_{2}A_{3}A_{4}}) \leq\frac{2}{\sqrt{18}}$.
From Fig.1, one gets that the
optimal lower bounds of $C_a(\rho_{A_1|A_2A_3})$ and $C_a(\rho_{A_1|A_2A_3A_4})$ are $0.249$ and $0.471$, respectively, attained at $x=2$.

\begin{figure}[htpb]
\renewcommand{\figurename}{Fig.}
\centering
\includegraphics[width=6.5cm]{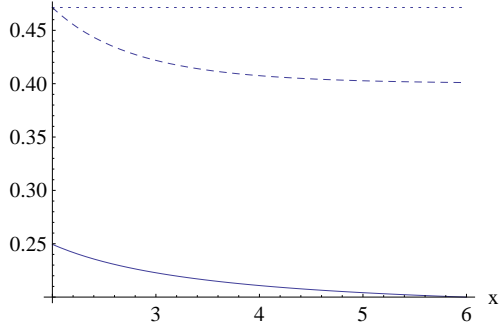}
\caption{{\small Solid line is the lower bound of $C_a(\rho_{A_1|A_2A_3})$,
dashed line is the lower bound of $C_a(\rho_{A_1|A_2A_3A_4})$ as functions of $x\geq2$,
and dotted line is the upper bound of $C_a(\rho_{A_1|A_2A_3})$ and $C_a(\rho_{A_1|A_2A_3A_4})$.}}
\label{Fig.1}
\end{figure}

\section{Monogamy of entanglement of formation}
The entanglement of formation of a pure state $|\psi\rangle\in H_A\otimes H_B$ is defined by
\begin{equation}\label{S}
E(|\psi\rangle)=S(\rho_A),
\end{equation}
where $\rho_A=Tr_{B}(|\psi\rangle\langle\psi|)$
and $S(\rho)=Tr(\rho\log_2\rho)$. For a bipartite
mixed state $\rho_{AB}\in H_A\otimes H_B$, the entanglement of formation is given by
\begin{equation}\label{EE}
E(\rho_{AB})=\min_{\{p_i,|\psi_i\rangle\}}\sum_ip_iE(|\psi_i\rangle),
\end{equation}
with the infimum taking over all possible decompositions of
$\rho_{AB}$ in a mixture of pure states
$\rho_{AB}=\sum_ip_i|\psi_i\rangle \langle \psi_i|$, where
$p_i\geq0$ and $\sum_ip_i=1$.

It has been shown that the entanglement of formation does not satisfy
the inequality $E_{AB}+E_{AC}\leq E_{A|BC}$ \cite{PRA61052306}.
Rather it satisfies \cite{024304},
\begin{equation}
E^\alpha_{A|B_1B_2...B_{N-1}}\geq E_{AB_1}^\alpha+E_{AB_2}^\alpha+...+E^\alpha_{AB_{N-1}},
\end{equation}
where $\alpha\geq\sqrt{2}.$

The corresponding
entanglement of assistance (EoA) \cite{OC} is defined
in terms of the entropy of entanglement \cite{Ea} for a tripartite pure
state $|\psi\rangle_{ABC}$,
\begin{equation}
E_a(|\psi\rangle_{ABC})\equiv E_a(\rho_{AB})=\max_{\{p_i,|\psi_i\rangle\}}\sum_ip_iE(|\psi_i\rangle),
\end{equation}
which is maximized over all possible decompositions of $\rho_{AB}=Tr_{C}(|\psi\rangle_{ABC})=\sum_ip_i|\psi_i\rangle \langle \psi_i|$,
with $p_i\geq0$ and $\sum_ip_i=1$.
For any $N$-qubit pure state $|\psi\rangle\in H_A
\otimes H_{B_1}\otimes...\otimes H_{B_{N-1}}$, it has been shown that the entanglement of assistance
satisfies \cite{024304},
\begin{equation}
E(|\psi\rangle_{A|B_1B_2...B_{N-1}})\leq
\sum_{i=1}^{N-1}E_a(\rho_{AB_i}).
\end{equation}

In fact, generally we can prove the following results for the $n$-qubit generalized W-class states
about the entanglement of formation and the entanglement of assistance.

\begin{theorem}\label{E}
For the $n$-qubit generalized W-class states $|\psi\rangle\in H_{A_1}\otimes H_{A_2}\otimes...\otimes H_{A_n}$, we have
\begin{eqnarray}\label{2}
E(|\psi\rangle_{A_1|A_2...A_{n}})\leq\sum_{i=2}^{{n}}E(\rho_{A_1A_{i}}),
\end{eqnarray}
where $\rho_{A_1A_i}$, $2\leq i\leq n$, is the $2$-qubit reduced density matrix of $|\psi\rangle$.
\end{theorem}

[Proof]
For the $n$-qubit generalized W-class states $|\psi\rangle$, we have
\begin{eqnarray}\nonumber
E(|\psi\rangle_{A_1|A_2...A_{n}})
&=&f\left(C^2(|\psi\rangle_{A_1|A_2...A_{n}})\right)\\[1mm]\nonumber
&=&f(\sum_{i=2}^{n}C^2(\rho_{A_1A_{i}}))\\[1mm]\nonumber
&\leq& \sum_{i=2}^{n}f(C^2(\rho_{A_1A_{i}}))\\[1mm]\nonumber
&=& \sum_{i=2}^{n}E(\rho_{A_1A_{i}}),
\end{eqnarray}
where for simplify, we have denoted $f(x)=h(\frac{1+\sqrt{1-x}}{2})$
with $h(x)=-x\log_2(x)-(1-x)\log_2(1-x).$
We have used in the first and last equalities that the entanglement
of formation obeys the relation $E(\rho)=f(C^2(\rho))$ for a bipartite $2\otimes D$, $D\geq2$, quantum state $\rho$
\cite{062343}. The second equality is due to the fact that
$C^2(|\psi\rangle_{A_1...A_{n}})=\sum_{i=2}^{n}C^2(\rho_{A_1A_{i}}).$
The inequality is due to the fact $f(x+y)\leq f(x)+f(y)$.
\qed

As for the entanglement of assistance, we have the following conclusion.
\begin{theorem}
For the $n$-qubit generalized W-class states $|\psi\rangle\in H_{A_1}\otimes H_{A_2}\otimes...\otimes H_{A_n}$, we have
\begin{eqnarray}\label{Eofa}
E(\rho_{A_1|A_{j_1}...A_{j_{m-1}}})\leq\sum_{i=1}^{{m-1}}E_a(\rho_{A_1A_{j_i}}),
\end{eqnarray}
where $\rho_{A_1|A_{j_1}...A_{j_{m-1}}}$ is the $m$-qubit reduced density matrix of
$|\psi\rangle$, $2\leq m\leq n$.
\end{theorem}

[Proof] From the lemma 2 of Ref.\cite{JSK1}, one has  $\rho_{A_1|A_{j_1}...A_{j_{m-1}}}$ of $|\psi\rangle$ is  a mixture of a  generalized $W$ class state and  vacuum.
Then, we have
\begin{eqnarray}\nonumber
E(\rho_{A_1|A_{j_1}...A_{j_{m-1}}})
&\leq&\sum_hp_hE(|\psi\rangle^h_{A_1|A_{j_1}...A_{j_{m-1}}})\\[1mm]\nonumber
&\leq&\sum_hp_h\sum_{i=1}^{m-1}E(\rho^h_{A_1A_{j_i}})\\[1mm]\nonumber
&=& \sum_{i=1}^{m-1}\left[\sum_hp_hE(\rho^h_{A_1A_{j_i}})\right]\\[1mm]\nonumber
&\leq&\sum_{i=1}^{m-1}\left[\sum_hp_h\left(\sum_jq_jE(|\psi_j\rangle^h_{A_1A_{j_i}}\langle\psi_j|)\right)\right]\\[1mm]\nonumber
&=&\sum_{i=1}^{m-1}\sum_{hj} p_hq_jE(|\psi_j\rangle^h_{A_1A_{j_i}}\langle\psi_j|).
\end{eqnarray}
We obtain the first inequality by noting that
$|\psi\rangle^h_{A_1|A_{j_1}...A_{j_{m-1}}}$ is  a generalized $W$ class state or vacuum\cite{JSK1}. When $|\psi\rangle^h_{A_1|A_{j_1}...A_{j_{m-1}}}$ is a generalized $W$ class state, then we have $E(|\psi\rangle^h_{A_1|A_{j_1}...A_{j_{m-1}}})\leq \sum_{i=1}^{m-1}E(\rho^h_{A_1A_{j_i}})$; When $|\psi\rangle^h_{A_1|A_{j_1}...A_{j_{m-1}}}$ is a vacuum, then we have $E(|\psi\rangle^h_{A_1|A_{j_1}...A_{j_{m-1}}})=0\leq \sum_{i=1}^{m-1}E(\rho^h_{A_1A_{j_i}})$.
The second inequality is due to the definition of the entanglement of formation (\ref{EE}) for mixed quantum states.
Since $\sum_{hj}p_hq_j=1$ and $\sum_{hj} p_hq_j|\psi_j\rangle^h_{A_1A_{j_i}}\langle\psi_j|$ is a pure
decomposition of $\rho_{A_1A_{j_i}}$, we have (\ref{Eofa}).
\qed

\section{Conclusions and remarks}
Entanglement monogamy is a fundamental property of multipartite
entangled states. We have shown the monogamy for the $x$-power of concurrence of assistance
$C_a(\rho_{A_1|A_{j_i}...A_{j_{m-1}}})$ of the $m$-qubit reduced density matrices, $2\leq m\leq n$,
for the $n$-qubit generalized $W-$class states.
The monogamy relations for the entanglement of formation and the entanglement of assistance
the monogamy relation for the $n$-qubit generalized W-class states have been also investigated.
These relations give rise to the restrictions of entanglement distribution among the qubits
in generalized $W-$class states.

\bigskip
\noindent{\bf Acknowledgments}\, \,
This work is supported by NSFC 11675113, 11605083 and 11275131.
Research Award Fund for
natural science foundation of Shandong province No.ZR2014AP013.

\end{document}